# Collaborative Filtering by Personality Diagnosis: A Hybrid Memory- and Model-Based Approach

**David M. Pennock**
NEC Research Institute
4 Independence Way
Princeton, NJ 08540
dpennock@research.nj.nec.com

**Eric Horvitz**
Microsoft Research
One Microsoft Way
Redmond, WA 98052-6399
horvitz@microsoft.com

**Steve Lawrence and C. Lee Giles**
NEC Research Institute
4 Independence Way
Princeton, NJ 08540
{lawrence,giles}@research.nj.nec.com

## Abstract

The growth of Internet commerce has stimulated the use of collaborative filtering (CF) algorithms as recommender systems. Such systems leverage knowledge about the known preferences of multiple users to recommend items of interest to other users. CF methods have been harnessed to make recommendations about such items as web pages, movies, books, and toys. Researchers have proposed and evaluated many approaches for generating recommendations. We describe and evaluate a new method called *personality diagnosis (PD)*. Given a user's preferences for some items, we compute the probability that he or she is of the same "personality type" as other users, and, in turn, the probability that he or she will like new items. PD retains some of the advantages of traditional similarity-weighting techniques in that all data is brought to bear on each prediction and new data can be added easily and incrementally. Additionally, PD has a meaningful probabilistic interpretation, which may be leveraged to justify, explain, and augment results. We report empirical results on the EachMovie database of movie ratings, and on user profile data collected from the CiteSeer digital library of Computer Science research papers. The probabilistic framework naturally supports a variety of descriptive measurements—in particular, we consider the applicability of a value of information (VOI) computation.

## 1 INTRODUCTION

The goal of collaborative filtering (CF) is to predict the preferences of one user, referred to as the active user, based on the preferences of a group of users. For example, given the active user's ratings for several movies and a database of other users' ratings, the system predicts how the active user would rate unseen movies. The key idea is that the active user will prefer those items that like-minded people prefer, or even that dissimilar people don't prefer. The effectiveness of any CF algorithm is ultimately predicated on the underlying assumption that human preferences are correlated—if they were not, then informed prediction would not be possible. There does not seem to be a single, obvious way to predict preferences, nor to evaluate effectiveness, and many different algorithms and evaluation criteria have been proposed and tested. Most comparisons to date have been empirical or qualitative in nature [Billsus and Pazzani, 1998; Breese et al., 1998; Konstan and Herlocker, 1997; Resnick and Varian, 1997; Resnick et al., 1994; Shardanand and Maes, 1995], though some worst-case performance bounds have been derived [Freund et al., 1998; Nakamura and Abe, 1998], some general principles advocated [Freund et al., 1998], and some fundamental limitations explicated [Pennock et al., 2000]. Initial methods were statistical, though several researchers have recently cast CF as a machine learning problem [Basu et al., 1998; Billsus and Pazzani, 1998; Nakamura and Abe 1998] or as a list-ranking problem [Cohen et al., 1999; Freund et al., 1998].

Breese et al. [1998] identify two major classes of prediction algorithms. Memory-based algorithms maintain a database of all users' known preferences for all items, and, for each prediction, perform some computation across the entire database. On the other hand, model-based algorithms first compile the users' preferences into a descriptive model of users, items, and/or ratings; recommendations are then generated by appealing to the model. Memory-based methods are simpler, seem to work reasonably well in practice, and new data can be added easily and incrementally. However, this approach can become computationally expensive, in terms of both time and space complexity, as the size of the database grows. Additionally, these methods generally cannot provide explanations of predictions or further insights into the data. For model-based algorithms, the model itself may offer added value beyond its predictive capabilities by highlighting certain correlations in the data, offering an intuitive rationale for recommendations, or simply making assumptions more explicit. Memory requirements for the model are generally less than for storing the full database. Predictions can be calculated quickly once the model is generated, though the time complexity to compile the data into a model may be prohibitive, and adding one new data point may require a full recompilation.

In this paper, we propose and evaluate a CF method called personality diagnosis (PD) that can be seen as a hybrid between memory- and model-based approaches. All data is maintained throughout the process, new data can be added incrementally, and predictions have meaningful probabilistic semantics. Each user's reported preferences are interpreted as a manifestation of their underlying *personality type*. For our purposes, personality type is encoded simply as a vector of the user's "true" ratings for titles in the database. It is assumed that users report ratings with Gaussian error. Given the active user's known ratings of items, we compute the probability that he or she has the same personality type as every other user, and then compute the probability that he or she will like some new item. The full details of the algorithm are given in Section 3.

PD retains some of the advantages of both memory- and model-based algorithms, namely simplicity, extensibility, normative grounding, and explanatory power. In Section 4, we evaluate PD's predictive accuracy on the EachMovie ratings data set, and on data gathered from the CiteSeer digital library's access logs. For large amounts of data, a straightforward application of PD suffers from the same time and space complexity concerns as memory-based methods. In Section 5, we describe how the probabilistic formalism naturally supports an expected value of information (VOI) computation. An interactive recommender could use VOI to favorably order queries for ratings, thereby mollifying what could otherwise be a tedious and frustrating process. VOI could also serve as a guide for pruning entries from the database with minimal loss of accuracy.

## 2    BACKGROUND AND NOTATION

Subsection 2.1 discusses previous research on collaborative filtering and recommender systems. Subsection 2.2 describes a general mathematical formulation of the CF problem and introduces any necessary notation.

### 2.1    COLLABORATIVE FILTERING APPROACHES

A variety of collaborative filters or recommender systems have been designed and deployed. The Tapestry system relied on each user to identify like-minded users manually [Goldberg et al., 1992]. GroupLens [Resnick et al., 1994] and Ringo [Shardanand and Maes, 1995], developed independently, were the first CF algorithms to automate prediction. Both are examples of the more general class of memory-based approaches, where for each prediction, some measure is calculated over the entire database of users' ratings. Typically, a similarity score between the active user and every other user is calculated. Predictions are generated by weighting each user's ratings proportionally to his or her similarity to the active user. A variety of similarity metrics are possible. Resnick et al. [1994] employ the Pearson correlation coefficient. Shardanand and Maes [1995] test a few metrics, including correlation and mean squared difference. Breese et al. [1998] propose the use of vector similarity, based on the vector cosine measure often employed in information retrieval. All of the memory-based algorithms cited predict the active user's rating as a similarity-weighted sum of the others users' ratings, though other combination methods, such as a weighted product, are equally plausible. Basu et al. [1998] explore the use of additional sources of information (for example, the age or sex of users, or the genre of movies) to aid prediction.

Breese et al. [1998] identify a second general class of CF algorithms called model-based algorithms. In this approach, an underlying model of user preferences is first constructed, from which predictions are inferred. The authors describe and evaluate two probabilistic models, which they term the Bayesian clustering and Bayesian network models. In the first model, like-minded users are clustered together into classes. Given his or her class membership, a user's ratings are assumed to be independent (i.e., the model structure is that of a naïve Bayesian network). The number of classes and the parameters of the model are learned from the data. The second model also employs a Bayesian network, but of a different form. Variables in the network are titles and their values are the allowable ratings. Both the structure of the network, which encodes the dependencies between titles, and the conditional probabilities are learned from the data. See [Breese et al., 1998] for the full description of these two models. Ungar and Foster [1998] also suggest clustering as a natural preprocessing step for CF. Both users and titles are classified into groups; for each category of users, the probability that they like each category of titles is estimated. The authors compare the results of several statistical techniques for clustering and model estimation, using both synthetic and real data.

CF technology is in current use in several Internet commerce applications [Schafer et al., 1999]. For example, the University of Minnesota's GroupLens and MovieLens[1] research projects spawned Net Perceptions,[2] a successful Internet startup offering personalization and recommendation services. Alexa[3] is a web browser plug-in that recommends related links based in part on other people's web surfing habits. A growing number of online retailers, including Amazon.com, CDNow.com, and Levis.com, employ CF methods to recommend items to their customers [Schafer et al., 1999]. CF tools originally developed at Microsoft Research are now included with the Commerce Edition of Microsoft's SiteServer,[4] and are currently in use at multiple sites.

---

[1] http://movielens.umn.edu/
[2] http://www.netperceptions.com/
[3] http://www.alexa.com/

## 2.2 FORMAL FRAMEWORK AND NOTATION

A CF algorithm recommends items or titles to the active user based on the ratings of other users. Let $n$ be the number of users, $T$ the set of all titles, and $m = |T|$ the total number of titles. Denote the $n \times m$ matrix of all users' ratings for all titles as $\mathbf{R}$. More specifically, the rating of user $i$ for title $j$ is $R_{ij}$, where each $R_{ij} \in \Re \cup \{\perp\}$ is either a real number or $\perp$, the symbol for "no rating". We overload notation slightly and use $\mathbf{R}_i$ to denote the $i$th row of $\mathbf{R}$, or the vector of all of user $i$'s ratings. Distinguish one user $a \in \{1, 2, \ldots, n\}$ as the active user. Define $NR \in T$ to be the subset of titles that the active user has not rated, and thus for which we would like to provide predictions. That is, title $j$ is in the set $NR$ if and only if $R_{aj} = \perp$.

In general terms, a collaborative filter is a function $f$ that takes as input all ratings for all users, and replaces some or all of the "no rating" symbols with predicted ratings. Call this new matrix $P$.

$$P_{aj} = \begin{cases} R_{aj} & : \text{ if } R_{aj} \neq \perp \\ f_a(\mathbf{R}) & : \text{ if } R_{aj} = \perp \end{cases}$$

# 3 COLLABORATIVE FILTERING BY PERSONALITY DIAGNOSIS

Traditional memory-based CF algorithms (e.g., similarity-weighted summations like GroupLens and Ringo) work reasonably well in practice, especially when the active user has rated a significant number of titles [Breese et al., 1998]. These algorithms are designed for, and evaluated on, predictive accuracy. Little else can be gleaned from their results, and the outcome of comparative experiments can depend to an unquantifiable extent on the chosen data set and/or evaluation criteria. In an effort to explore more semantically meaningful approaches, we propose a simple model of how people rate titles, and describe an associated personality diagnosis (PD) algorithm to generate predictions. One benefit of this approach is that the modeling assumptions are made explicit and are thus amenable to scrutiny, modification, and even empirical validation.

Our model posits that user $i$'s personality type can be described as a vector $\mathbf{R}_i^{\text{true}} = \langle R_{i1}^{\text{true}}, R_{i2}^{\text{true}}, \ldots, R_{im}^{\text{true}} \rangle$ of

[4]http://www.microsoft.com/DirectAccess/ products/sscommerce

"true" ratings for all seen titles. These encode his or her underlying, internal preferences for titles, and are not directly accessible by the designer of a CF system. We assume that users report ratings for titles they've seen with Gaussian noise. That is, user $i$'s reported rating for title $j$ is drawn from an independent normal distribution with mean $R_{ij}^{\text{true}}$. Specifically,

$$\Pr(R_{ij} = x | R_{ij}^{\text{true}} = y) \propto e^{-(x-y)^2 / 2\sigma^2}, \quad (1)$$

where $\sigma$ is a free parameter. Thus the same user may report different ratings on different occasions, perhaps depending on the context of any other titles rated in the same session, on his or her mood, or on other external factors. All factors are summarized here as Gaussian noise. Given the user's personality type, his or her ratings are assumed independent. (If $y = \perp$ in Equation 1, then we assign a uniform distribution over ratings.)

We further assume that the distribution of personality types or ratings vectors in the database is representative of the distribution of personalities in the target population of users. That is, the prior probability $\Pr(\mathbf{R}_a^{\text{true}} = v)$ that the active user rates items according to a vector $v$ is given by the frequency that other users rate according to $v$. Instead of explicitly counting occurrences, we simply define $\mathbf{R}_a^{\text{true}}$ to be a random variable that can take on one of $n$ values—$\mathbf{R}_1, \mathbf{R}_2, \ldots, \mathbf{R}_n$—each with probability $1/n$.

$$\Pr(\mathbf{R}_a^{\text{true}} = \mathbf{R}_i) = \frac{1}{n} \quad (2)$$

From Equations 1 and 2, and given the active user's ratings, we can compute the probability that the active user is of the same personality type as any other user, by applying Bayes' rule.

$$\begin{aligned} &\Pr(\mathbf{R}_a^{\text{true}} = \mathbf{R}_i | R_{a1} = x_1, \ldots, R_{am} = x_m) \\ &\propto \quad \Pr(R_{a1} = x_1 | R_{a1}^{\text{true}} = R_{i1}) \\ &\cdots \Pr(R_{am} = x_m | R_{am}^{\text{true}} = R_{im}) \\ &\cdot \Pr(\mathbf{R}_a^{\text{true}} = \mathbf{R}_i) \end{aligned} \quad (3)$$

Once we compute this quantity for each user $i$, we can compute a probability distribution for the active user's rating of an unseen title $j$.

$$\begin{aligned} &\Pr(R_{aj} = x_j | R_{a1} = x_1, \ldots, R_{am} = x_m) \\ &= \sum_{i=1}^{n} \Pr(R_{aj} = x_j | \mathbf{R}_a^{\text{true}} = \mathbf{R}_i) \\ &\cdot \Pr(\mathbf{R}_a^{\text{true}} = \mathbf{R}_i | R_{a1} = x_1, \ldots, R_{am} = x_m) (4) \end{aligned}$$

where $j \in NR$. The algorithm has time and space complexity $O(nm)$, as do the memory-based methods described in Section 2.1. The model is depicted as a naïve Bayesian network in Figure 1. It has the same structure as a classical diagnostic model, and indeed the analogy is apt.

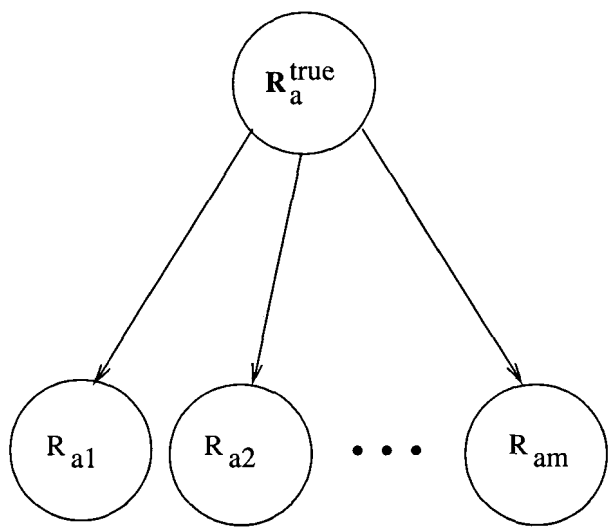

Figure 1: Naïve Bayesian network semantics for the PD model. Actual ratings are independent and normally distributed given the underlying "true" personality type.

Table 1: Average absolute deviation scores on the Each-Movie data for PD and the four algorithms tested in Breese et al. [1998]. Lower scores are better. Correlation and vector similarity are memory-based algorithms; Bayesian clustering and Bayesian network are model-based. PD performed best under all conditions. Bayesian clustering and Bayesian network scores are transcribed directly from Breese et al. [1998].

| Algorithm | Protocol | | | |
|-----------|----------|---------|--------|--------|
|           | All But 1 | Given 10 | Given 5 | Given 2 |
| PD        | 0.965    | 0.986   | 1.016  | 1.040  |
| Correl.   | 0.999    | 1.069   | 1.145  | 1.296  |
| V. Sim.   | 1.000    | 1.029   | 1.073  | 1.114  |
| B. Clust. | 1.103    | 1.138   | 1.144  | 1.127  |
| B. Net.   | 1.066    | 1.139   | 1.154  | 1.143  |

We observe ratings ("symptoms") and compute the probability that each personality type ("disease") is the cause using Equation 3. We can then compute the probability of rating values for an unseen title $j$ using Equation 4. We return the *most probable* rating as our prediction.

An alternative but equivalent interpretation of this model is as follows. The active user is assumed to be "generated" by choosing one of the other users uniformly at random and adding Gaussian noise to his or her ratings. Given the active user's known ratings, we infer the probability that he or she is actually one of the other users, and then compute the probabilities for ratings of other items. PD can also be thought of as a clustering method [Breese et al., 1998; Ungar and Foster, 1998] with exactly one user per cluster. The general approach of casting CF as a classification problem has been advocated and examined previously [Basu et al., 1998; Billsus and Pazzani, 1998; Cohen et al., 1999; Freund et al., 1998; Nakamura and Abe 1998]. Note that in the PD model, the only free parameter is $\sigma$.

## 4    EMPIRICAL RESULTS

We have evaluated the PD algorithm on two datasets: a subset of the EachMovie database, available from the Digital Equipment Research Center, and user profile data from the CiteSeer digital library of Computer Science research papers.

### 4.1    EACHMOVIE

The EachMovie data contains many thousands of users' ratings for various movies, elicited on a scale from 0 to 5. We used the same subset of the data as Breese et al. [1998], consisting of 1623 titles, 381,862 ratings, 5000 users in the training set, and 4119 users in the test set. On average, each user rated about 46.3 movies. To carry out testing, we withhold some of the ratings of users in the test set and attempt to predict them using a CF algorithm. Again following the methodology of Breese et al. [1998], we employ four different protocols. Under the first protocol, called *all but one*, we withhold for prediction only one rating for each user in the test set; all other ratings are used as input for the CF algorithm. In the other three protocols, *given ten*, *given five*, and *given two*, we retain the given number of ratings for each user for input to the algorithm, and try to predict the rest. Each protocol admits less information than the previous, and we should expect a corresponding decrease in accuracy. If a user does not rate enough movies to satisfy a particular protocol, then he or she is dropped from that experiment. The parameter $\sigma$ was set initially to correspond with the variance in the ratings data, and then hand-tuned during the training phase to a value of 2.5.

Breese et al. [1998] propose two evaluation criteria to measure accuracy: rank scoring and average absolute deviation. We consider here only the latter. Let $p$ be the total number of predictions made for all users in the test set. Then the average absolute deviation is simply $1/p \sum |P_{aj} - R_{aj}|$.

The results are summarized in Table 1. Scores for the two model-based methods—Bayesian clustering and Bayesian network—are transcribed directly from Breese et al. [1998]; we did not replicate these experiments. We did however reimplement and retest the memory-based algorithms of correlation and vector similarity. Our results for correlation match fairly well with those reported in Breese et al. [1998], though vector similarity performed much better in our tests. We are not sure of the source of the discrepancy. Due to randomization, we almost certainly did not withhold exactly the same titles for prediction as those authors. PD performed better than each of the other four algorithms under all four protocols. In fact, PD under the given-ten protocol outperformed correlation under the all-but-one protocol, which was the previous best score. Note that, among the other four algorithms, none was a strict winner.

Table 2: Average absolute deviation scores for PD, correlation, and vector similarity on the EachMovie data for extreme ratings 0.5 above or 0.5 below the overall average rating.

| Algorithm | Protocol | | | |
|---|---|---|---|---|
| | All But 1 | Given 10 | Given 5 | Given 2 |
| PD | 1.030 | 1.087 | 1.129 | 1.163 |
| Correl. | 1.130 | 1.211 | 1.282 | 1.424 |
| V. Sim. | 1.108 | 1.127 | 1.167 | 1.189 |

Table 3: Significance levels of the differences in scores between PD and correlation, and between PD and vector similarity, computed using the randomization test on Each-Movie data. Low significance levels indicate that differences in results are unlikely to be coincidental.

| | PD vs. Correl. | PD vs. V. Sim. |
|---|---|---|
| All But 1 | 0.0842 | 0.0803 |
| All But 1 (extreme) | $4 \times 10^{-5}$ | $1.01 \times 10^{-3}$ |
| Given 10 | $10^{-5}$ | $2 \times 10^{-5}$ |
| Given 10 (extreme) | $10^{-5}$ | $3.4 \times 10^{-4}$ |
| Given 5 | $10^{-5}$ | $10^{-5}$ |
| Given 5 (extreme) | $10^{-5}$ | $6.1 \times 10^{-4}$ |
| Given 2 | $10^{-5}$ | $10^{-5}$ |
| Given 2 (extreme) | $10^{-5}$ | 0.0335 |

Shardanand and Maes [1995] argue that CF accuracy is most crucial when predicting *extreme* (very high or very low) ratings for titles. Intuitively, since the end goal is typically to provide recommendations or warnings, high accuracy on the best and worst titles is most important, while poor performance on mediocre titles is acceptable. Table 2 displays the average absolute deviation of PD, correlation, and vector similarity for which the true score is 0.5 above the average or 0.5 below the average. In other words, deviations were computed only when the withheld rating $R_{ij}$ was less than $\bar{R} - 0.5$ or greater than $\bar{R} + 0.5$, where $\bar{R}$ is the overall average rating in the data set. PD outperformed correlation and vector similarity when predicting these extreme ratings.

Table 3 summarizes the statistical significance of the EachMovie results. We employed the randomization test [Fisher, 1966; Noreen, 1989] to compute *significance levels* for the differences in absolute average deviation between PD and correlation, and between PD and vector similarity. We proceeded by randomly shuffling together the deviation scores of the two competing algorithms in 100,000 different permutations, and, for each permutation, computing the difference in average absolute deviation between the two (now randomized) subsets. Table 3 reports the probability that a random difference is greater than or equal to the true difference. Low numbers indicate that the reported differences between algorithms are unlikely to be coincidental. These low significance levels effectively rule of the null hypothesis that the algorithms' deviation scores arise from the same distribution.

## 4.2 CITESEER

CiteSeer creates digital libraries of scientific literature [Lawrence et al., 1999]. A CiteSeer service that indexes computer science literature is available at http://csindex.com/, and currently indexes about 270,000 articles. CiteSeer uses explicit and implicit feedback in order to maintain user profiles that are used to recommend new documents to users [Bollacker et al., 1999]. The system logs a number of user actions that we use to create ratings for each document. User actions include viewing the details of a document, downloading a document, viewing related documents, and explicitly adding a document to a user's profile. We assigned a weight to each of the actions, as shown in Table 4, and computed a rating for each user–document pair as the sum of the respective weights for all actions that the user performed on the specific document (rounded to integers), resulting in a range of ratings from 0 to 6. The weights were chosen to correspond roughly to our intuition of what actions are most indicative of user preferences; we did not attempt to optimize the weights through any formal procedure.

The raw CitesSeer data is sparse; most users share documents with only one or two others users, and users must share at least two documents for any meaningful testing of the memory-based algorithms. Thus for the purpose of these experiments, we only included documents that were rated by 15 or more users (1575 documents), and we only included users that rated 2 or more of these popular documents (8244 users). There were a total of 32,736 ratings, or 3.97 ratings per user on average. The users were split into two equal subsets for training and testing. As more users access CiteSeer, and as we increase the amount of user profile information recorded, we expect the ratings matrix to fill in.

Results for PD, correlation, and vector similarity are sum-marized in Table 5. Due to the limited number of titles rated by each user, there was a reasonable amount of data only for the *all but one* and *given two* protocols. PD's predictions resulted in the smallest average absolute deviation under both protocols. Table 6 displays the three algorithms' average absolute deviation on "extreme" titles, for which the true score is 0.5 above the average or 0.5 below the average. Again, PD outperformed the two memory-based methods that we implemented. Table 7 reports the statistical significance of these results. These comparisons suggest that it is very unlikely that the reported differences between PD and the other two algorithms are spurious.

Table 4: User actions for documents in CiteSeer, along with the weights assigned to each action.

| Action | Weight |
|---|---|
| Add document to profile | 2 |
| Download document | 1 |
| View document details | 0.5 |
| View bibliography | 0.5 |
| View page image | 0.5 |
| Ignore recommendation | -1.0 |
| View documents from the same source | 0.5 |
| View document overlap | 1 |
| Correct document details | 1 |
| View citation context | 1 |
| View related documents | 0.5 |

Table 5: Average absolute deviation scores for PD, correlation, and vector similarity on the CiteSeer data.

| Algorithm | Protocol | |
|---|---|---|
| | All But 1 | Given 2 |
| PD | 0.562 | 0.589 |
| Correl. | 0.708 | 0.795 |
| V. Sim. | 0.647 | 0.668 |

## 5 HARNESSING VALUE OF INFORMATION IN RECOMMENDER SYSTEMS

Formulating collaborative filtering as the diagnosis of personality under uncertainty provides opportunities for leveraging information- and decision-theoretic methods to provide functionalities beyond the core prediction service. We have been exploring the use of the expected value of information (VOI) in conjunction with CF. VOI computation identifies, via a cost-benefit analysis, the most valuable new information to acquire in the context of a current probability distribution over states of interest [Howard, 1968]. In the current context, a VOI analysis can be used to drive a hypothetico-deductive cycle [Horvitz et al., 1988] that identifies at each step the most valuable ratings information to seek next from a user, so as to maximize the quality of recommendations.

Table 6: Average absolute deviation scores for PD, correlation, and vector similarity on the CiteSeer data for extreme ratings 0.5 above or 0.5 below the overall average rating.

| Algorithm | Protocol | |
|---|---|---|
| | All But 1 | Given 2 |
| PD | 0.535 | 0.606 |
| Correl. | 0.912 | 0.957 |
| V. Sim. | 1.084 | 1.111 |

Table 7: Significance levels of the differences in scores between PD and correlation, and between PD and vector similarity, on CiteSeer data. Low numbers indicate high confidence that the differences are real.

| | PD vs. Correl. | PD vs. V. Sim. |
|---|---|---|
| All But 1 | $3.7 \times 10^{-4}$ | 0.114 |
| All But 1 (extreme) | $10^{-5}$ | $10^{-5}$ |
| Given 2 | $10^{-5}$ | 0.101 |
| Given 2 (extreme) | $10^{-5}$ | $10^{-5}$ |

Recommender systems in real-world applications have been designed to acquire information by explicitly asking users to rate a set of titles or by implicitly watching the browsing or purchasing behavior of users. Employing a VOI analysis makes feasible an optional service that could be used in an initial phase of information gathering or in an ongoing manner as an adjunct to implicit observation of a user's interests. VOI-based queries can minimize the number of explicit ratings asked of users while maximizing the accuracy of the personality diagnosis. The use of general formulations of expected value of information as well as simpler information-theoretic approximations to VOI hold opportunity for endowing recommender systems with intelligence about evidence gathering. Information-theoretic approximations employ measures of the expected change in the information content with observation, such as relative entropy [Bassat, 1978]. Such methods have been used with success in several Bayesian diagnostic systems [Heckerman et al., 1992].

Building a VOI service requires the added specification of utility functions that captures the cost of querying a user for his or her ratings. A reasonable class of utility models includes functions that cast cost as a monotonic function of the number of items that a user has been asked to evaluate. Such models reflect the increasing frustration that users may have with each additional rating task. In an explicit service guided by such a cost function, users are queried about titles in decreasing VOI order, until the expected cost of additional requests outweighs the expected benefit of improved accuracy.

Beyond the use of VOI to guide the gathering of preference information, we are pursuing the offline use of VOI to compress the amount of data required to produce good recommendations. We can compute the average information gain of titles and/or users in the data set and eliminate those of low value accordingly. Such an approach can provide a means for both alleviating memory requirements and improving the running time of recommender systems with as little impact on accuracy as possible.

# 6 CONCLUSION

We have described a new algorithm for collaborative filtering (CF) called personality diagnosis (PD), which can be thought of as a hybrid between existing memory- and model-based algorithms. Like memory-based methods, PD is fairly straightforward, maintains all data, and does not require a compilation step to incorporate new data. Most memory-based algorithms operate as a "black box": efficacy is evaluated by examining only the accuracy of the output. Since results do not have a meaningful interpretation, the reason for success or failure is often hard to explain, and the search for improvements becomes largely a trial-and-error process. The PD algorithm is based on a simple and reasonable probabilistic model of how people rate titles. Like other model-based approaches, its assumptions are explicit, and its results have a meaningful probabilistic interpretation. According to absolute deviation, on movie ratings data, PD makes better predictions than four other algorithms—two memory-based and two model-based—under four conditions of varying information about the active user. PD also outperforms the two memory-based algorithms on a subset of research paper ratings data extracted from the CiteSeer digital library. Most results are highly statistically significant. Finally, we discussed how value of information might be used in the context of an interactive CF algorithm or a data compression scheme.

We plan to extend the PD framework to incorporate user and title information beyond ratings—for example, user age groups or movie genres. We will also consider relaxing some of the assumptions of the model, for example allowing some conditional dependence among ratings, or letting $\sigma$ vary across users. Future empirical tests will evaluate PD against other CF algorithms, on additional data sets, and according to other proposed evaluation metrics, including Breese et al.'s *rank scoring* criteria [1998]. Perhaps the ultimate gauge for CF algorithms, though, is user satisfaction. We plan to employ PD and other CF algorithms to recommend research papers to CiteSeer users and investigate which algorithms generate the highest click-through rates.

## Acknowledgments

Thanks to Jack Breese, Carl Kadie, Frans Coetzee, and the anonymous reviewers for suggestions, advice, comments, and pointers to related work.